\documentclass[%
 reprint,
superscriptaddress,
 amsmath,amssymb,
 aps,pra
]{revtex4-2}

\usepackage{graphicx}
\usepackage{dcolumn}
\usepackage{bm}

\usepackage{subfigure}
\usepackage{xcolor}
\usepackage{amsmath}
\usepackage{physics}
\usepackage{float}
\begin{document}

\preprint{APS/123-QED}
\title{Trapping multiple absorbing particles in air using an optical fiber by photophoretic forces}

\author{Souvik Sil}
\author{Anita Pahi}
\author{Aman Anil Punse}
\author{Ayan Banerjee}

\affiliation{Department of Physical Sciences, Indian Institute of Science Education and Research Kolkata, Kalyani - 741246, India}
\email{Corresponding author: ayan@iiserkol.ac.in}






\begin{abstract}
We demonstrate photophoretic force-based optical trapping of multiple absorbing particles in air by loosely focusing a Gaussian beam with a series of convex lenses of different focal lengths, and investigate the dependence of the number of trapped particles and their sizes on the focal length. We observe the formation of particle chains at a particular focal length, and measure the dynamic range of optical trapping for each lens system. We then develop a numerical simulation to explain the observed dynamic range of trapping by estimating the temperature distribution across a particle surface, and determining the photophoretic force. Our simulation results are in reasonable agreement with experimental results. Interestingly, we also observe that the average size of trapped particles reduces as we increase the lens focal lengths, which suggests that intensity gradients may somehow be involved in the mechanism of photophoretic trapping.  
\end{abstract}


\maketitle

The photophoretic force - being  almost five order magnitude higher compared to radiation pressure force - can easily overcome the effects of gravity, and has therefore facilitated robust optical traps in air \cite{desyatnikov2009,shvedov2010giant,zhang2011trapping}. However, trapping multiple particles controllably remains a challenge. Recently, the use of structured light fields, i.e. light containing regions of alternating dark and bright zones have been proposed towards achieving this. Such complex light fields may be obtained using holographic optical bottle beams \cite{alpmann2012holographic}, vortex beams \cite{shvedov_2009} , volume speckle fields \cite{shvedov2010}, tapered rings \cite{liu_2014_photophoretic} or even optical lattices \cite{shvedov2012optical}, which have thus been shown to be able to trap multiple particles in air.  Recently,  Zhang et al. \cite{zhang2012observation} trapped multiple particles using a focused fundamental Gaussian ($TEM_{00}$) beam - however, they did not describe the trapping mechanism in detail. An important task thus remains to trap multiple particles using a simple Gaussian beam, and vary its intensity to study its effect on the trapping of multiple particles.

In this paper, we address this problem by experimentally demonstrating a configuration where a simple Gaussian beam propagating against gravity and used in combination with six plano-convex lenses of different focal lengths (25 - 150 mm), is used to trap a linear chain of absorbing printer toner particles. We determine the performance of each lens in multiple particle trapping, and observe that the lens with 75 mm focal length leads to the longest chain of particles in our system. We are able to confine particles of different sizes along the trapping beam, and measure the $z$-position of the trapped particles. We then estimate the intensity of the laser at the location of the particles, and also determine the particle sizes using image-processing tools. We proceed to calculate the photophoretic $\Delta T$ force acting on the trapped particles using numerical simulations, and estimate the dynamic range of the photophoretic trap for a specific size distribution for each convex lens. Our simulation results are in resonable agreement with the experimentally obtained values - but also raise important and intriguing questions about the detailed mechanism of photophoretic trapping.

\begin{figure}[htbp]
\centering
\includegraphics[width=0.4\textwidth]{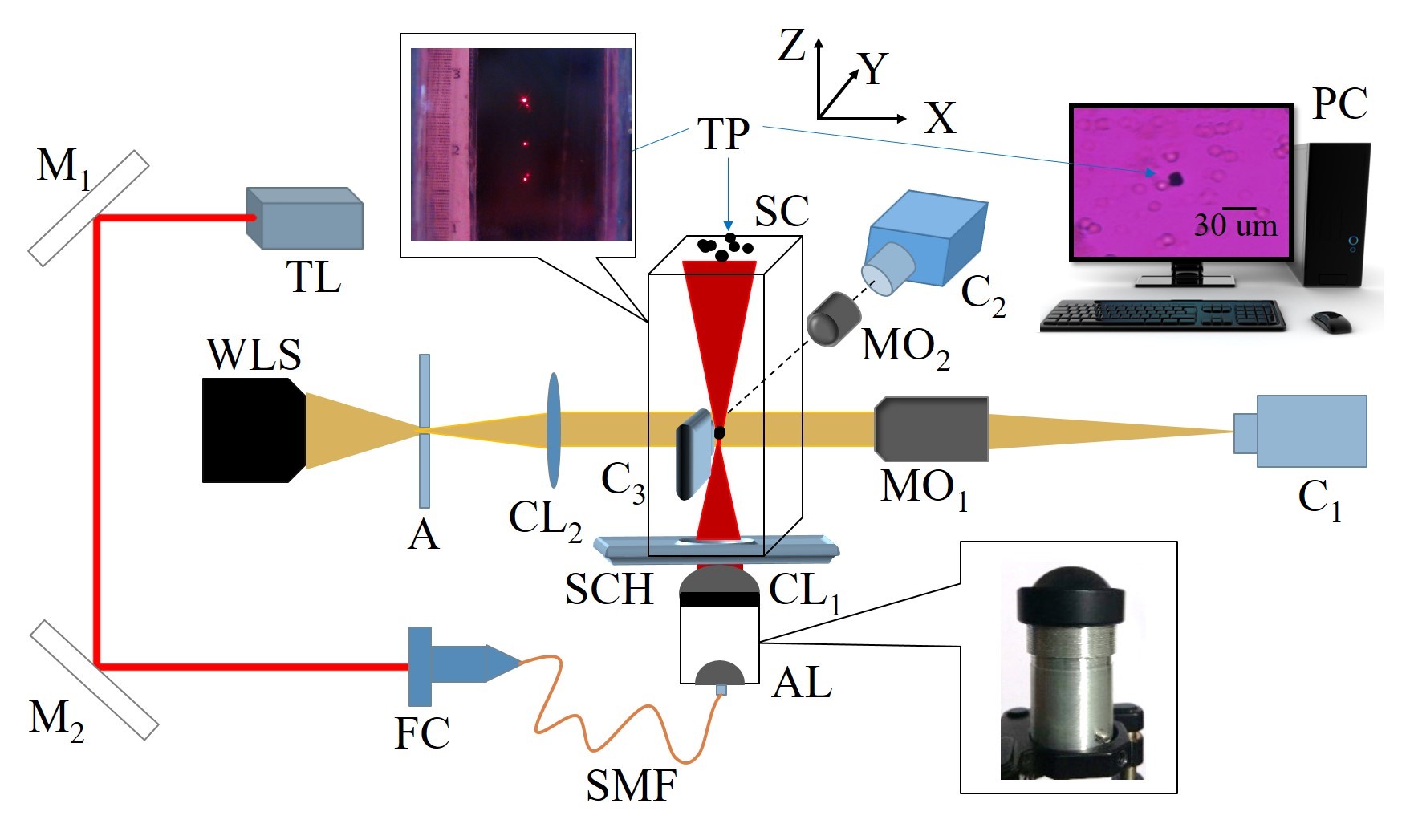}
\caption{Schematic of the experimental setup; TL: Trapping laser; M: Mirrors; FC: Fiber coupler; SMF: Single-mode fiber; AL: Aspheric Lens; CL: Convex lens; SCH: Sample Chamber Holder; SC: Sample Chamber; WLS: White Light Source; A: Apperture; C: Cameras; MO: Microscope 10x objective; PC: Computer. Left top inset: zoomed-in image of the sample chamber with three trapped particles. Right top inset: Single trapped particle image using $C_3$. Bottom-right inset: Indigenous collimator with AL for collimating the SMF output beam, with $CL_1$ used for further focusing.}
\label{Fig1}
\end{figure}

A schematic of the experimental setup (see Ref.~\cite{sil_2017} for more details) is shown in \ref{Fig1}, where a single-mode (SM) optical fiber (SM-600 to 800 nm) is coupled to a laser source of 640 nm wavelength with a maximum power of 300 mW. The output beam from the fiber is collimated by our home built collimator using an aspheric lens (AL in Fig.~\ref{Fig1}) whose numerical aperture (NA) matches with the numerical aperture (NA) of the fiber, and a convex lens (CL$_1$ in Fig.~\ref{Fig1}) is attached in the same mount to focus the beam for trapping (see inset of Fig.~\ref{Fig1}). We use printer toner particles which are trapped inside our home built glass sample chamber, and are imaged (see top right and left insets of Fig.~\ref{Fig1}) on cameras (C$_1$ and C$_2$ with the help of 10x collection objectives (MO$_1$) and (MO$_2$) along the x and y axes of the trapping plane  using a white light source that is collimated by a convex lens (CL$_2$).  A third camera C$_3$ is used to determine the axial position of a trapped particle in the z-direction by taking an image of the trapped particle along with a measuring scale affixed to the sample chamber. The axial trapping positions of the trapped particles are measured from a pixel by pixel analysis of the corresponding images taken by camera C$_3$ with around 20 $\mu m$ accuracy.
\begin{figure*}
\centering
\fbox{\includegraphics[width=0.63\textwidth]{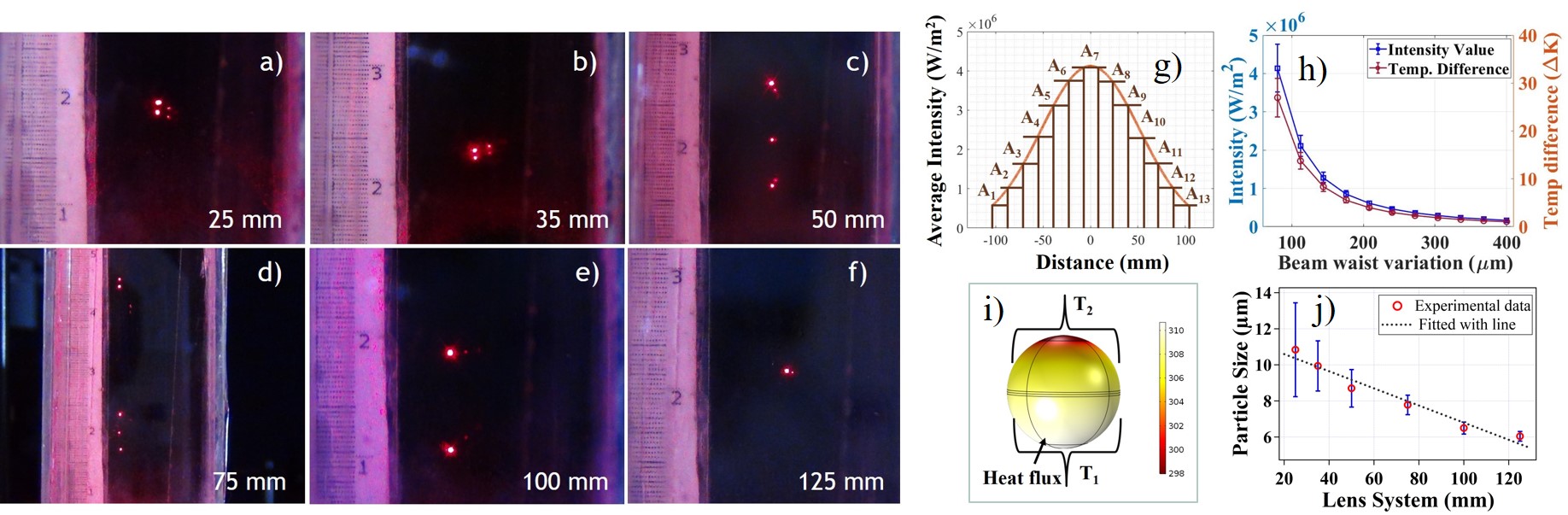}}
\caption{Particles trapped for different trapping convex lenses of focal length f = a) 25 mm b) 35 mm; c) 50 mm; d) 75 mm; e) 100 mm; f) 125 mm; red dots depict the scattered light from trapped particles; \textbf{g)} Schematic for the measurement of average intensity for a certain beam waist value; \textbf{h)} Variation of average laser intensity (left axis: solid blue line with open squares) and temperature difference $\Delta T$ across the particle surface (right axis: solid red line with open circles)  with the different beam waist values ; \textbf{i)} Temperature distribution across the particle surface, with heat flux given over the lower hemisphere, and $T_1$ and $T_2$ being the average temperatures of the lower and upper hemispheres, respectively; \textbf{j)} Variation of the average radii of trapped particles  for different convex lenses.}
\label{Fig2}
\end{figure*}
For trapping particles, we use a series of different convex lenses (CL$_1$) of focal length range from f = 25 mm to 125 mm. For each case, we locate the z position and subsequently take the images of the trapped particle for particle size and mass determination. Interestingly, we observe that the  number of particles trapped are different for different convex lenses, which are shown in Fig.~\ref{Fig2} a) to f). We trap around 10-15 particles for each lens to develop our statistics - other than the focal lengths of 100 and 125 mm where the trapping proved to challenging and we could trap only 5 and 2 particles, respectively. We perform 3 experiments of 15 minutes for each lens to ensure uniformity in experimental conditions. The number of trapped particles in a single experimental run also increases with increasing focal length (see Table \ref{tab:Table 1 Dynamic Range}, with a maximum of 5 particles trapped in a chain at a focal length of 75 mm (Fig.~\ref{Fig2} d)) -  the particles being stably trapped both below and above the lens focus. Beyond this focal length, the number of trapped particles decreases.  Moreover, the particles are trapped at different z positions for various convex lenses, which signifies that the equilibrium positions of trapped particles are different for each lens.  The experimental results of the dynamic range for different convex lenses are displayed in Table  \ref{tab:Table 1 Dynamic Range}. From the Table, it is also clear that the distance away from the focus where particles are trapped increases with increasing focal length. We do not observe any trapping at focal lengths higher than 125 mm in our experimental system. The particle sizes at each focal length are determined using the methodology described in Ref.~\cite{sil2020study}.

\begin{table*}[htbb]
\centering
\caption{\label{tab:Table 1 Dynamic Range}
 \bf Experimental and simulation results of dynamic range for different focal lengths of convex lenses. The 1$\sigma$ errors in the experimentally measured values are around 0.04 mm.}
\begin{tabular}{ccccc}
\hline
Convex Lens system  & Experimental & Beam Waist  &  Simulation & Maximum number of \\ Focal length (mm) & Dynamic range (mm) & ($\mu m$) & Dynamic range (mm) & trapped particles (N)  \\&&&& in a single run\\
\hline
25 & 0.82 - 6.27 & 61.4 - 404.7 & 2.02 - 5.50  & 2  \\
35 & 3.03 - 8.32 & 125.5 - 344.8 & 3.51 - 8.97 & 2  \\
50 & 3.90 - 8.63 & 110.9 - 245.2 & 5.01 - 12.82 & 3  \\
75 & 7.16 - 14.31 & 134.2 - 267.6 & 7.51 - 19.23 & 5  \\
100 & 14.70 - 22.54 & 211.4 - 323.6 & 9.99 - 25.63 & 2  \\
125 & 20.48 - 30.44 & 235.2 - 349.1 & 12.45 - 32.02 & 1  \\
\hline
\end{tabular}
\end{table*}
We now attempt to develop a numerical simulation to determine whether the known theory of photophoretic forces account for the axial trapping range we experimentally observe for lenses of different focal lengths.  The photophoretic force is based on two types of forces - typically referred to in the literature as F$_{\Delta T}$ and F$_{\Delta \alpha}$ \cite{horvath_2014}. Particles are confined in the axial direction due to F$_{\Delta T}$ which balances gravity ($F_g$), while radially, a restoring force is generated by the helical motion of particles caused by the transverse photophoretic body force (F$_{\Delta \alpha}$), which applies a torque on particles due to its interaction with gravity \cite{sil_2017,sil2020study}. Since it is basically the photophoretic $\Delta T$ force that balances gravity to cause axial trapping (or levitation), we determine this force using the multiphysics simulation software COMSOL. Thus, we determine the temperature difference across a particle surface from the value of its thermal conductivity and the intensity of the laser incident on the surface, and calculate the $\Delta T$ force using the semi-empirical formula provided by Rohatschek \cite{rohatschek_1995}, according to which, 
\begin{equation}
 F = D \ \frac{p^*}{p} a \ \Delta T   
 \label{deltaT}
\end{equation}
Where, D is a constant determined entirely by the state of the gas, $p^*$ is characteristic pressure which depends on the particle radius (a), thermal creep coefficient ($\kappa$), and the average velocity of air molecules.  We now compare the calculated value of this force with the gravitational force, and determine the axial position range over which these two forces balance each other. Note that here we neglect the radiation pressure force ($F_r$) as for absorbing particles, the value of ($F_r$) is much lesser than the gravitational force on the particle.

For determining the intensity incident on the particle surface along the axial direction ($z$), we need to find out the beam waist $w_z$ as a function of $z$  for different convex lenses. For that, we first measure the input beam waist $w_{0f}$ for a lens of focal length $f$ using the well known moving edge technique, where a sharp edge is moved across the transverse cross-section of the beam and the intensity profile thereby built up, is fit with the Gaussian error function \cite{de2009measurement}. Then, by inserting the respective values of $w_{0f}$ and $z_{0f}$ for a convex lens of focal length f into the Gaussian beam propagation equation $w(z) = w_{0f}\sqrt{\left(1+\frac{z^2}{z_0^2}\right)}$, we are able to measure the beam waist $W(z)$ at any $z$ position. Thus, from the experimentally determined axial trapping ranges (Table \ref{tab:Table 1 Dynamic Range}), we conclude that the corresponding beam waists vary from 146 $\mu m$ to 330 $\mu m$. Accordingly, in the first step of our simulation, we numerically vary the beam waist from 16 $\mu m$ to 400 $\mu m$ with a step size of 32 $\mu m$, and correspondingly measure the average intensity at each beam waist value. Here, we define the average intensity for a specific beam waist value as the laser intensity experienced by a particle of 16 $\mu m$ diameter (since the average diameter of our sample particles is 16 $\mu m$). To measure the average intensity experienced by a particle, we plot a 1-D Gaussian curve with a spot size of 2$w$, and divide it into segments of width 16 $\mu m$, extending from $-w$ to $+w$ as shown in Fig.~\ref{Fig2} g). Thus, for a beam waist of 112 $\mu m$, we obtain 13 such segments (A$_1$ to A$_{13}$). Since we maintain the laser power at 70 mW throughout the experiment, the area under the Gaussian intensity distribution  ($A$) should equal 70 mW (P). Then, the average power $<p_i>$ for each segment $i$ is $<p_i> = \dfrac{<a>P}{A}$, where $<a> = \Sigma_{i=1}^{N}\dfrac{A_i}{N}$, with $A_i$ the area of segment $i$, and $N$ the total number of segments. Finally, the average intensity ($<I>$) for a specific beam waist ($w$) is obtained as $<I> = \dfrac{<p>}{d}$, where $d$ is the particle diameter. We repeat this process for other beam waist values, with the results shown in Fig.~\ref{Fig2} h) (solid blue line in the left axis). We also calculate the standard deviation (SD) for the average intensities of each beam waist, and find out that lower the beam waist value, the larger the SD of $<I>$. The next step is to measure temperature distribution across the particle surface using the COMSOL Heat transfer model.

In the COMSOL simulation, we assume the particle to be spherical, and we use the `Heat transfer in solids (time-dependent)' model to find the temperature distribution across the particle whose diameter is again 16 $\mu m$. We choose the incident heat flux ($H$)  as $H = \chi <I>$, where $\chi$ is the absorptivity of the particle. We consider this as a spatially distributed heat source on the particle surface [see Fig.~\ref{Fig2} i)], which is introduced at the lower hemisphere, and set the ambient temperature of the particle at 298 K. The thermal conductivity, density, specific heat of the printer toner particle we trap are provided as user-defined values. A representative temperature distribution across the particle surface for a certain heat flux value is shown in Fig.~\ref{Fig2} i).  We note down the temperature distributions generated by COMSOL for the beam waists corresponding to all lenses we use in our experiments. Now, we measure the temperature difference $(\Delta T)$ across the particle by taking the difference between the average temperatures of the lower $(T_1)$ and upper $(T_2)$ hemispheres [see Fig.~\ref{Fig2} i)]. The variation of the temperature difference ($\Delta T$) for different beam waists are shown in Fig.~\ref{Fig2} h) (red solid line in the right axis). Note that in Fig.~\ref{Fig2} h), we discard the initial two points for beam waists 16 $\mu m$ and 48 $\mu m$ since, at these two points, the average intensities are quite large and $\Delta T$ is around 677 K and 75 K, respectively, which exceeds/is very close to the melting temperature of toner particles (which is 80K above room temperature). Now, the variation of temperature difference $\Delta T$ in Fig.~\ref{Fig2} h) is obviously similar to the average intensity variation with beam waist [Fig.~\ref{Fig2} h) (solid blue line)], which is understandable, since the temperature difference is linearly dependent on the laser intensity obtained from the beam waist. 

\begin{figure}[htbp]
\centering
\fbox{\includegraphics[width=0.85\linewidth]{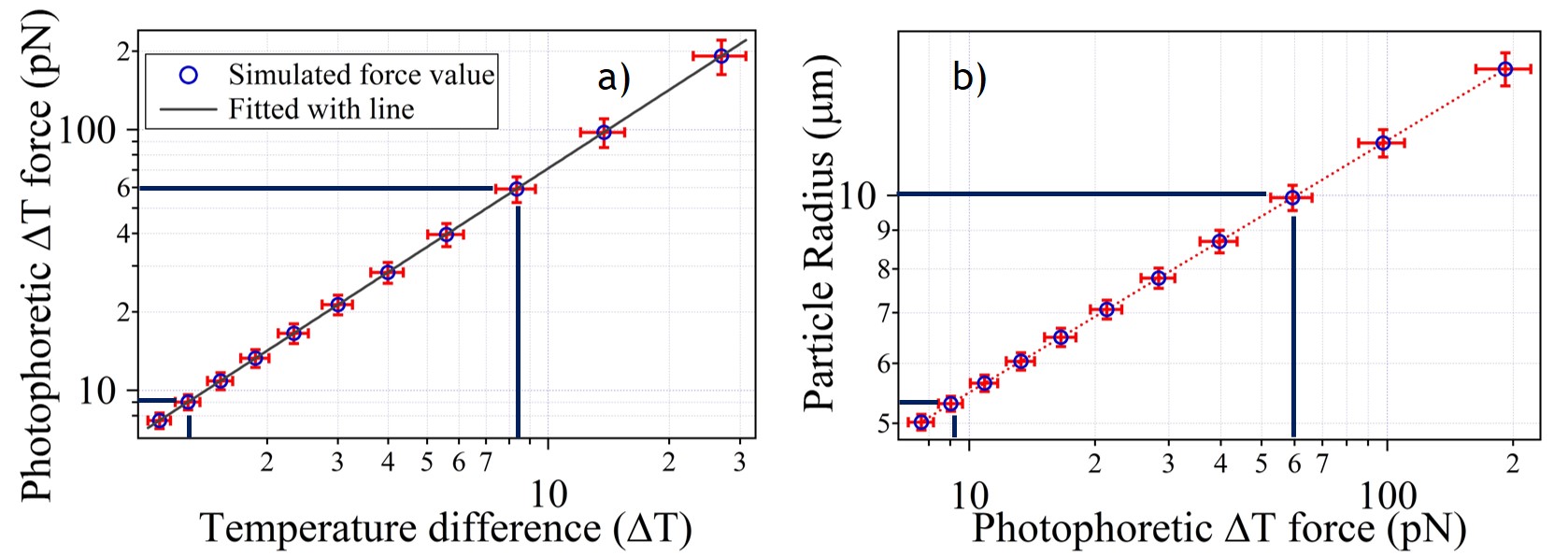}} 
\caption{a)Variation of photophoretic $\Delta T$ force with the temperature difference of the particle ($\Delta T$); blue circular points are simulated force values fitted with a line (black) b) Possible trapped particle size ($\mu m$) depending upon $F_{\Delta T}$ and $F_g$. Blue solid lines in both graphs depict the region where particles can be trapped.}
\label{Fig3}
\end{figure}

Thus, from the simulation, we obtain the  values of $\Delta T$ for a particle for different beam waist values i.e. different axial planes. Then, using Eq.~ \ref{deltaT}, we determine the photophoretic (F$_{\Delta T}$) force, and display its variation with $\Delta T$ in Fig.~\ref{Fig3} a). Obviously, F$_{\Delta T}$ has a linear dependence on $\Delta T$, but the significance of this graph is to displaythe axial range where values of F$_{\Delta T}$ will balance the weight of the particle $F_g$.  The simulation also reveals that $\Delta T$ for a particle is not too sensitively dependent on the particle diameter (changing only by 11 \% when we change the radius by 1 $\mu m$) for the same laser intensity, so that - assuming a constant value of F$_{\Delta T}$, and then using the force balance condition, we are able to calculate the radii of the particles which may be axially levitated. The variation of particle radius $\mu m$ with F$_{\Delta T}$ force is shown in Fig.~\ref{Fig3} b). Note that the analysis of the particle size from SEM images comes around $6.2 \pm 1.5~\mu m$, but experimentally we determine the average size of trapped particles to be  $8.1 \pm 2.5~ \mu m$. For this specific range, the corresponding $\Delta T$ and $F_{\Delta T}$ should vary between 1.27- 8.34K and 9 - 60 pN, respectively. The solid blue lines in Fig.~\ref{Fig3}  depict the regions of the above-mentioned range. 

The axial trapping range from the simulation comes out to be 144 - 368 $\mu m$, which is close to our experimentally determined range (146 - 330 $\mu m$). We now compare the simulation values for each lens with the corresponding experimental results in  Table \ref{tab:Table 1 Dynamic Range}. Understandably, the axial trapping range increases with increasing focal length, since the Rayleigh range increases as the focal length is increased, so that the intensity increases slowly with $z$. The predicted ranges from the simulation are within 5-40\% of those observed in experiments, other than the lowest and highest focal lengths (25 and 125 mm). For the 25 mm case, the large error in the determination of the average intensity itself leads to similar error in the determination of both $\Delta T$ and F$_{\Delta T}$, which may explain our experimental observation that particles of larger size (12 $\mu m$ diameter, more than 3$\sigma$ away from the higher range in the particle size distribution) are observed to be trapped in this case. It is also possible that we have particles agglomerating while they are falling in the air or inside the trap itself \cite{bera_2016}. For the higher focal lengths (100 and 125 mm), we trap very few particles, and inadequate statistics may be the reason of the deviation of around 40\% from the numerically estimated axial distance closer to the focus. Finally, the simulation also suggests that particles should be trapped for 150 mm and 200 mm focal length lenses with axial ranges 14.9 - 38.4 mm and 19.7 - 51.1 mm (not shown in Table). However, we do not observe any trapping for those lenses in our experiments.

A very interesting observation we have in our experiments is that the average size of particles clearly reduces with increasing focal length of lenses (Fig.~\ref{Fig2} j). This is not what is suggested by the simulations, which are all performed for the same particle diameter. The simulation also does not throw any light on why we observe the maximum number of particles at the focal length of 75 mm. Here, an interesting observation we have is that while the average radius of particles from the SEM image analysis is of $6.3\pm 1.5~ \mu m$, the average radius we obtain for the 75 mm focal length case is $7.2\pm1 0.7~\mu m$ - which lie within 1$\sigma$ of each other. Thus, it is possible that the 75 mm focal length lens provides the most optimum trapping conditions for our sample particles so that we get the longest chain of particles here. However, this still does not explain why we observe very few particles trapped for the lenses of even longer focal length. Here, our hypothesis is that - while the same values of the light intensity is achieved for all the lenses at different $z$ positions - what most definitely changes is the intensity gradient, that clearly decreases axially as we increase the focal length. Thus, we hypothesize that the axial intensity gradient is optimum for the 75 mm focal length lens for our size distribution of particles, and is too high or too low for focal lengths higher or lower than this particular value. Now, the present theory of photophoretic trapping does not include any role for the intensity gradient in determining trapping conditions - but our experiments seem to suggest that this theory needs to be revisited and a more complete theory developed to explain our experimental observations. It is also important to emphasize that radial trapping - which we do not study here - may also play a role in explaining our experimental data.

In conclusion, we experimentally demonstrate that multiple absorbing particles can be trapped using a single, very loosely focused Gaussian beam. We trap particles using a series of convex lenses starting from a focal length of 25 to 125 mm, and find the largest number of particles in a single experimental run in the case of the 75 mm focal length lens. Trapping becomes increasingly difficult for higher focal length lenses, and we do not observe trapping beyond a focal length of 125 mm. The average size of trapped particles also decreases as we increase lens focal lengths. We experimentally measure the axial dynamic range of trapping for each convex lens, and attempt to validate it using numerical simulations employing COMSOL. Thus, we determine the temperature difference across a trapped particle, and calculate the photophoretic force that would balance the gravitational force employing the the extant theory for photophoretic forces. We obtain a reasonable agreement in the prediction of the dynamic range. We would like to emphasize, however, that these results are only for our input beam size, and while our prescription of determining the axial dynamic range of trapping would remain valid, actual results would certainly change. However - our model does not explain why we observe lower sized particles being trapped for progressively higher focal lengths, why a chain of particles is trapped for a particular focal length, or  why no particles are trapped beyond a focal length of 125 mm. We do observe that the initial size distribution of our particles match the average size of particles trapped using the 75 mm lens within experimental error. To explain our overall observations, we hypothesise that the intensity gradient of light, previously not invoked to explain photophoretic trapping, actually may play a role in determining the optimum conditions that facilitate such trapping and levitation. Clearly the theory of photophoretic trapping needs to be revisited, and more experiments need to be performed to verify our observations. We look forward eagerly to such developments.

The authors acknowledge IISER Kolkata, an autonomous institution funded by the Ministry of Education (MoE), Govt of India for funding and laboratory space. SS thanks CSIR, MoE for fellowship support.


\end{document}